\documentclass[draft,nofootinbib,pre,twocolumn,showpacs,showkeys,groupaddress,preprintnumbers,floatfix]{revtex4-1}

\usepackage{dynlearn}
\usepackage{physics}

\usepackage{mathrsfs}
\usepackage[english]{babel}
\usepackage{amssymb,amscd}

\newcommand{\kB}{k_\text{B}}

\newcommand{\MV}[1]{\left\langle #1 \right\rangle}

\newcommand{\intP}{\int_{\mathcal{P}(\mathcal{H})} \!\!\!\!\!\!\!\!\!}

\newcommand{\PH}{\mathcal{P}(\mathcal{H})}
\newcommand{\CP}[1]{\mathbb{C}P^{#1}}

\def\tbf #1 {\textbf{#1} }

\begin{document}

\def\ourTitle{%
Geometric Quantum Thermodynamics
}

\def\ourAbstract{%
Building on parallels between geometric quantum mechanics and classical
mechanics, we explore an alternative basis for quantum thermodynamics that
exploits the differential geometry of the underlying state space. We focus on
microcanonical and canonical ensembles, looking at the geometric counterpart of
Gibbs ensembles for distributions on the space of quantum states. We show that
one can define quantum heat and work in an intrinsic way, including
single-trajectory work. We reformulate thermodynamic entropy in a way that
accords with classical, quantum, and information-theoretic entropies. We give
both the First and Second Laws of Thermodynamics and Jarzynki's Fluctuation
Theorem. Overall, this results in a more transparent physics than
conventionally available. The mathematical structure and physical intuitions
underlying classical and quantum dynamics are seen to be closely aligned. The
experimental relevance is brought out via a stochastic model for chiral
molecules (in the two-state approximation) and Josephson junctions.
Numerically, we demonstrate this invariably leads to the emergence of the
geometric canonical ensemble.
}

\def\ourKeywords{%
quantum thermodynamics, differential geometry, classical thermodynamics,
statistical mechanics
}

\hypersetup{
  pdfauthor={Fabio Anza},
  pdftitle={\ourTitle},
  pdfsubject={\ourAbstract},
  pdfkeywords={\ourKeywords},
  pdfproducer={},
  pdfcreator={}
}


\title{\ourTitle}

\author{Fabio Anza}
\email{fanza@ucdavis.edu}

\author{James P. Crutchfield}
\email{chaos@ucdavis.edu}

\affiliation{Complexity Sciences Center and Physics Department,
University of California at Davis, One Shields Avenue, Davis, CA 95616}

\date{\today}
\bibliographystyle{unsrt}

\begin{abstract}
\ourAbstract
\end{abstract}



\preprint{\arxiv{2008.08683}}

\date{\today}
\maketitle





\section{Introduction}
\label{sec:Intro}

Geometric quantum mechanics (GQM) exploits the tools of differential geometry
to analyze the phenomenology of quantum systems. It does so by focusing on the
interplay between statistics and geometry of quantum state space.

For finite-dimensional quantum systems---that we consider here---the state
space $\mathcal{H}$ is isomorphic to a complex projective space $\mathbb{C}P^n$
of dimension $n=D-1$, where $D\coloneqq \mathrm{dim}~\mathcal{H}$. Our goal is
to explore the statistical and thermodynamic consequences of the geometric
approach. In particular, structural and informational properties can be
properly formulated. And, the close parallels in the mathematical foundations
of classical and quantum dynamics become clear.

To the best of our knowledge, the development of the geometric formalisms
started with early insights from Strocchi \cite{STROCCHI1966} and then work by
Kibble \cite{Kibble1979}, Marsden \cite{Chern76a}, Heslot \cite{Heslot1985},
Gibbons \cite{Gibbons1992}, Ashtekar and Shilling
\cite{Ashtekar1995,Ashtekar1999}, and a host of others
\cite{Hugh95,Brody2001,Bengtsson2017,Carinena2007,Chruscinski2006,Marmo2010,Avron2020,Pastorello2015,Pastorello2015a,Pastorello2016,Clemente-Gallardo2013}.
Although geometric tools for quantum mechanics are an interesting topic in
their own right, the following explores their consequences for statistical
mechanics and nonequilibrium thermodynamics.

As one example in this direction, Brody and Hughston
\cite{Brody1998,Brody2007,Brody2016} showed that a statistical mechanics
treatment of quantum systems based on the geometric formulation differs from
standard quantum statistical mechanics: The former can describe phase
transitions away from the thermodynamic limit, the latter not
\cite{Brody2007b}. This arises, most directly, since the geometric formulation
puts quantum mechanics on the same footing as the classical mechanics of phase
space \cite{Heslot1985,STROCCHI1966}, bringing to light the symplectic geometry
of quantum state space. It is then straightforward to build on the principles
of classical statistical mechanics to layout a version of quantum statistical
mechanics that takes advantage of such state-space features.

That said, these insights do not come for free. The conundrum of a consistent
foundation of thermodynamic behavior arises. On the one hand, we have quantum
statistical mechanics---a description of macroscopic behavior that, despite
limitations, has proven to be remarkably successful. On the other,
transitioning from microphysics to macrophysics via quantum mechanics is
conceptually different than via classical mechanics. Consistency between these
approaches begs for a conceptually unique route from microphysics to
macrophysics.

With this broad perspective in mind, unifying the two coexisting statistical
mechanics of quantum systems, though challenging, deserves further attention.
To address the challenge, the following advocates a geometric development of a
practical, macroscopic companion of geometric quantum statistical mechanics---a
{geometric quantum thermodynamics}.

Beyond foundations, geometric quantum thermodynamics is all the more timely due
to recent success in driving thermodynamics down to the mesoscopic scale. There
statistical fluctuations, quantum fluctuations, and collective behavior not
only cannot be neglected, but are essential. Largely, this push is articulated
in two research thrusts: \emph{stochastic thermodynamics}
\cite{Seifert2008,Seifert2012} and \emph{quantum thermodynamics}
\cite{Gemmer2004,Deffner2019}. The following draws ideas and tools from both,
in effect showing that geometric tools provide a robust and
conceptually-incisive crossover between them.

Our development unfolds as follows. First, it recalls the basic elements of
geometric quantum mechanics. Second, it shows how this formalism emerges
naturally in a thermodynamic context. Third, it describes our version of the
statistical treatment of geometric quantum mechanics---what we refer to as
geometric quantum statistical mechanics. Fourth, it builds on this to establish
two fundamental equations of geometric quantum thermodynamics. The first is a
novel version of the first law of quantum thermodynamics, with its definition
of quantum heat and quantum work. The second is a quantum version of
Jarzynski's inequality---one that does not require a two-time measurement
scheme. Fifth, the development proposes an experiment that highlights geometric
quantum thermodynamics' practical relevance. Finally, it expands on the
geometric approach's increasing relevance to the thermodynamics of quantum
information and computing.

\section{Geometric Quantum Mechanics}
\label{sec:GQM}

Geometric quantum mechanics arose from efforts to exploit differential geometry
to probe the often-counterintuitive behaviors of quantum systems. This section
summarizes the relevant concepts, adapting them to our needs. Detailed
expositions are found in the original literature
\cite{STROCCHI1966,Kibble1979,Chern76a,Hugh95,Heslot1985,Gibbons1992,Ashtekar1995,Ashtekar1999,Brody2001,Bengtsson2017,Carinena2007,Chruscinski2006,Marmo2010,Avron2020,Pastorello2015,Pastorello2015a,Pastorello2016,Clemente-Gallardo2013}. Here, we present the main ideas in a constructive way, focusing on the aspects
that are of direct relevance to thermodynamic behavior.

Any statistical mechanics requires an appropriate, workable concept of
ensemble. To do this, one identifies ensembles with coordinate-invariant
measures on the space of quantum states, a treatment first introduced in Ref.
\cite{Brody1998}. We call these distributions \emph{geometric quantum states}
and in Ref. \cite{Anza20a} we give a generic procedure to compute them in a
quantum thermodynamic setting of a small system interacting with a large
environment.

Achieving this, though, requires a series of technical steps. The first
identifies the manifold of pure states and defines their observables. The
second introduces a suitable metric, scalar product, and coordinate-invariant
volume element for the pure-state manifold. From these, the third step derives
the evolution operator and equations of motion. Finally, states are described
via functionals that map observables to scalar values. This is done so that the
associated ensembles are coordinate-invariant measures.

Our quantum system of interest has Hilbert space $\mathcal{H}$ of finite
dimension $D$. The space of pure states is therefore the \emph{complex projective
space} $\mathcal{P}\left( \mathcal{H} \right) \sim \mathbb{C}P^{D-1}$
\cite{Bengtsson2017}.  Given an arbitrary basis $\left\{\ket{e_\alpha}
\right\}_{\alpha=0}^{D-1}$ a generic pure state is parametrized by $D$
complex homogeneous coordinates $Z^\alpha$, up to normalization and an overall
phase:
\begin{align*}
\ket{\psi} = \sum_{\alpha=0}^{D-1} Z^\alpha \ket{e_\alpha}
  ~,
\end{align*}
where $Z \in \mathbb{C}^{D}$, $Z \sim \lambda Z$, and $\lambda \in
\mathbb{C}/\left\{ 0\right\}$.

For example, the pure state $Z_{\mathrm{qubit}}$ of a single qubit can be given
real coordinates: $Z_{\mathrm{qubit}} = (\sqrt{p},\sqrt{1-p} e^{i\nu})$. An
observable $\mathcal{O}$ is a quadratic real function of the state. It
associates to each point of the pure-state manifold $\mathcal{P}\left(
\mathcal{H} \right)$ the expectation value $\bra{\psi} \mathcal{O} \ket{\psi}$
of the corresponding operator $\mathcal{O}$ on that state:
\begin{align}
\mathcal{O}(Z) = \sum_{\alpha,\beta} \mathcal{O}_{\alpha,\beta}Z^\alpha \overline{Z}^\beta
\label{eq:GQM_Observable}
\end{align}
and $\mathcal{O}_{\beta,\alpha} = \overline{\mathcal{O}}_{\alpha,\beta}$. And
so, $\mathcal{O}(Z) \in \mathbb{R}$.

These complex projective spaces are \emph{Kahler spaces}. This means there
is a function $K$, which in our case is $K = \log Z \cdot \overline{Z}$,
from which one obtains both a metric $g$:
\begin{align*}
g_{\alpha \overline{\beta}} = \frac{1}{2} \partial_\alpha \overline{\partial}_{\beta} \log Z \cdot \overline{Z}
  ~,
\end{align*}
with $\overline{g_{\alpha \beta} } = g_{\beta \alpha}$, and a symplectic
two-form:
\begin{align*}
\Omega = 2ig_{\alpha \beta} dZ^\alpha \wedge d\overline{Z}^\beta
  ~,
\end{align*}
using shorthand $\overline{\partial}_\alpha \coloneqq \partial/\partial
\overline{Z}^\alpha$. It is not too hard to see that these two structures are
parts of the Hermitian form that defines the scalar product
$\braket{\psi_1}{\psi_2}$ in $\mathcal{H}$. Indeed, using the standard notation,
one has \cite{Gibbons1992}:
\begin{align*}
\braket{\psi_1}{\psi_2} = g(Z_1,Z_2) + i \Omega(Z_1,Z_2)
  ~,
\end{align*}
Each geometric term provides an independent volume element.

Agreement between these volumes, together with invariance under unitary
transformations, selects a unique coordinate-invariant volume element $d
V_{FS}$ \cite{Brody1998}, based on the Fubini-Study metric on
$\mathbb{C}P^{D-1}$:
\begin{subequations}
\begin{align}
d V_{FS}
  & = \frac{1}{(D-1)!} \left( \frac{\Omega}{2}\right)
  \wedge \left( \frac{\Omega}{2}\right)
  \wedge \ldots \wedge
  \left( \frac{\Omega}{2}\right) \\
  & = \sqrt{\mathrm{det}\,g(Z,\overline{Z})} dZ d\overline{Z}
  ~.
\end{align}
\label{eq:FSMeasure}
\end{subequations}
(See also Ref. \cite{Bengtsson2017} for a textbook treatment.) Equipped with this unique volume element, the total volume of the pure-state manifold
$\mathbb{C}P^{D-1}$ is \cite{Bengtsson2017,Gibbons1992}:
\begin{align*}
\mathrm{Vol} \left( \mathbb{C}P^n\right) = \frac{\pi^{D-1}}{(D-1)!}
  ~.
\end{align*}

Since symplectic geometry is the correct environment in which to formulate
classical mechanics, one can see how the geometric formalism brings classical
and quantum mechanics closer together---a point previously raised by Strocchi
\cite{STROCCHI1966} and made particularly clear by Heslot \cite{Heslot1985}.
Indeed, as in classical mechanics, the symplectic two-form $\Omega$ is an
antisymmetric tensor with two indices that provides Poisson brackets,
Hamiltonian vector fields, and the respective dynamical evolution.

Given two functions $A$ and $B$ on manifold $\mathcal{P}(\mathcal{H})$ we have:
\begin{align*}
\Omega(A,B) & = \partial_\alpha A
  \overline{\partial}_\beta B \Omega^{\alpha \beta} \\
  & = \left\{ A,B\right\}
  ~,
\end{align*}
where we used $\Omega = \frac{1}{2} \Omega_{\alpha \beta} dZ^\alpha \wedge
d\overline{Z}^\beta$ and $\Omega^{\alpha \beta} = (\Omega^{-1})_{\alpha \beta}$
is the inverse: $\Omega^{\alpha \gamma} \Omega_{\gamma \beta} =
\delta^\alpha_\beta$. Using the symplectic two-form one can show that
Schr\"odinger's unitary evolution under operator $H$ is generated by a Killing
vector field $V_H$ as follows:
\begin{subequations}
\begin{align}
V_H^\alpha & = \Omega^{\alpha \beta} \partial_\beta h(Z) \\
\frac{dF}{dt} & = \left\{F,h \right\} 
\end{align}
\label{eq:LiouvilleDynamics}
\end{subequations}
where $h(Z) = \sum_{\alpha \beta}H_{\alpha \beta} Z^\alpha \overline{Z}^\beta$
and $F : \mathcal{P}(\mathcal{H}) \to \mathbb{R}$ is a real but otherwise
arbitrary function. Indeed, it can be shown that Schr\"odinger's equation is
nothing other than Hamilton's equations of motion in disguise
\cite{Bengtsson2017,Heslot1985}:
\begin{align}
\frac{d \ket{\psi_t}}{dt} = -i H \ket{\psi_t} \quad \Longleftrightarrow \quad  \frac{dF}{dt} = \left\{ F, h\right\}
  ~,
\label{eq:unitary_evolution}
\end{align}
for all $F$. Here, we use units in which $\hbar = 1$. 

This framework naturally views a quantum system's states as the functional
encoding that associates expectation values with observables; as done in the
$C^{*}$-algebra formulation of quantum mechanics \cite{Strocchi2008a}. Thus,
states are described via functionals $P[\mathcal{O}]$ from the algebra
$\mathcal{A}$ of observables to the reals: 
\begin{align*}
P[\mathcal{O}] = \int_{\mathcal{P}(\mathcal{H})}
  p(Z) \mathcal{O}(Z) dV_{FS} \in \mathbb{R}
  ~,
\end{align*}
for $p(Z) \geq 0$ and all $\mathcal{O} \in \mathcal{A}$. Here, $p$ is the
distribution associated to the functional $P$. It is important to note here
that $dV_{FS}$ and $\mathcal{O}(Z)$ are both invariant under coordinate
changes. Thus, for $P[\mathcal{O}]$ to be a scalar, $p(Z)$ must be a scalar
itself. A pure state $\ket{\psi} \in \mathcal{H}$ is represented by a
Dirac-delta functional concentrated on a single point of
$\mathcal{P}(\mathcal{H})$. However, Dirac delta-functions $\delta(\cdot)$ are
not invariant under coordinate changes: They transform with the inverse of the
Jacobian: $\delta \to \delta / \mathrm{det} J $.

To build an invariant quantity, then, we divide it by the square root
$\sqrt{g}$ of the metric's determinant. This transforms in the same way,
making their ratio $\tilde{\delta} = \delta/\sqrt{g}$ an invariant quantity.
This is a standard rescaling that turns coordinate-dependent measures, such as
Cartesian measure, into coordinate-invariant ones. And, this is how the
Fubini-Study measure Eq. (\ref{eq:FSMeasure}) is defined from the Cartesian
product measure. Thus:
\begin{align}
P_{\psi_0}[\mathcal{O}]
  & = \intP \tilde{\delta} [Z-Z_0] \nonumber
  \mathcal{O}(Z) dV_{FS}
  \label{eq:GQM_PureState} \\
  & = \mathcal{O}(Z_0) \nonumber \\
  & = \bra{\psi_0}\mathcal{O} \ket{\psi_0}
  ~,
\end{align}
where:
\begin{align*}
\tilde{\delta}[Z-Z_0]
  = \frac{1}{\sqrt{g}} \prod_{\alpha} \delta(Z^\alpha - Z_0^\alpha)
\end{align*}
and:
\begin{align*}
\delta(Z^\alpha - Z_0^\alpha)
  = \delta(\mathrm{Re}[Z^\alpha]
  - \mathrm{Re}[Z^\alpha_0])\delta(\mathrm{Im}[Z^\alpha]
  - \mathrm{Im}[Z^\alpha_0])
  ~.
\end{align*}
This extends by linearity to ensembles $\rho =
\sum_{k=1}^M p_k \ket{\psi_k} \bra{\psi_k}$ as:
\begin{align*}
P_\rho[\mathcal{O}] 
  & = \sum_{h=1}^M p_k  \intP \tilde{\delta} [Z-Z_k] \mathcal{O}(Z) dV_{FS} \\
  & = \sum_{h=1}^M p_k \mathcal{O}(Z_k) \\
  & = \sum_{h=1}^M p_k\bra{\psi_k}\mathcal{O}\ket{\psi_k} 
  ~.
\end{align*}

It is now quite natural to consider generalized ensembles that correspond to
functionals with a continuous measure on the pure-state manifold. Such ensembles have 
appeared previously in Refs. \cite{Brody2001,Brody2007,Brody1998,Brody2016} and elsewhere, 
where aspects of their properties have been investigated extensively. For our purposes, it will
be useful to look at them from the following point of view.

Consider a probability measure on the natural numbers: $\left\{p_k\right\}$
such that $p_k \geq 0$ and $\sum_k p_k = 1$. Now let $Z_k$ be a countable
collection of points in $\mathcal{P}(\mathcal{H})$, then $\delta_k(dZ)$ is the
Dirac measure concentrated on the point $Z_k$. Then, given $\left\{
p_k\right\}$ one can define the measure $\mu(dZ)$ on $\mathcal{P}(\mathcal{H})$
as:
\begin{align}
\mu(dZ) = \sum_{k=1}^{\infty} p_k \delta_k(dZ) \label{eq:countable}
  ~,
\end{align}
which gives precise meaning to the notion of a geometric quantum state with
support on a countably-infinite number of points. Indeed, with the measure in
Eq.(\ref{eq:countable}) and arbitrary observable function $\mathcal{O}(Z)$ one
has that:
\begin{align*}
P_{\infty}[\mathcal{O} ]
  & = \int_{\PH} \!\!\!\!\!\!\!\!\! \mathcal{O}(Z) \mu(dZ) \\
  & = \sum_{k=1}^{\infty} p_k \mathcal{O}(Z_k) 
  ~.
\end{align*}
In more general terms, calling $\mathcal{B}$ the Borel $\sigma$-algebra of the
open sets of $\PH$, then, this procedure defines a measure $\mu$ on $\PH$ such
that for a set $S \in \mathcal{B}$ one has:
\begin{align*}
\mu(S) & = \int_{S} \mu(dZ) \\
  & =  \sum_{k=1}^{\infty} p_k I(Z_k \in S)
  ~,
\end{align*}
where $I(Z_k \in S)$ is the indicator function which is $1$ if $Z_k \in S$ and
zero otherwise.

The resulting geometric quantum state has all the properties desired of an
appropriately-generalized pure-state ensemble: It preserves normalization and
convexity of linear combinations, each of its elements are invariant under
coordinate changes, and the entire functional $P_\infty$ is also invariant
under unitary transformations. With some abuse of language, we will often refer
to both the functional $P$ and their underlying measure $\mu$ as geometric
quantum states.

\section{Geometric quantum state and the thermodynamic limit}
\label{sec:Extension}

We are now equipped to address how the geometric formalism arises quite
naturally for subsystems of a larger system in a pure state; in particular, in
a quantum thermodynamic setting.

If we have a bipartite system $\mathcal{H}_{AB } = \mathcal{H}_A \otimes
\mathcal{H}_B$ and $\ket{\psi_{AB}} = \sum_{\alpha,i} \psi^{\alpha i}_{AB}
\ket{a_\alpha} \ket{b_i} \in \mathcal{H}_{AB}$, the partial trace over the
subsystem $B$ is:
\begin{align*}
\rho^A = \sum_{\alpha,\beta = 1}^{d_A}
  \rho^A_{\alpha\beta}  \ketbra{a_\alpha}{a_\beta}
  ~,
\end{align*}
where:
\begin{align*}
\rho_{\alpha\beta}^A
  & = \sum_{i=1}^{d_B} \psi^{\alpha i} \overline{\psi}^{\beta i} \\
  & =  (\psi \psi^\dagger)_{\alpha \beta}
  ~.
\end{align*}
$d_A$ and $d_B$ are $A$'s and $B$'s dimensions, respectively.
Hence, we can write the partial trace as:
\begin{align*}
\rho^A = \sum_{j=1}^{d_B} \ketbra{v_j}{v_j}
  ~,
\end{align*}
with $\ket{v_i} \in \mathcal{H}_A$ given as:
\begin{align*}
\ket{v_i} \coloneqq \sum_{\alpha=1}^{d_A} \psi^{\alpha i} \ket{a_\alpha}
  ~.
\end{align*}

However, $\ket{v_j}$ is not normalized. To address this, we notice that:
\begin{align*}
\braket{v_j}{v_k} & = (\psi^\dagger \psi)_{jk} \\
  & = \rho^B_{jk} \\
  & = \bra{b_j} \rho^B \ket{b_k}
  ~.
\end{align*}
This gives:
\begin{align*}
p_k^B & = \rho^B_{kk} \\
  & = \sum_{\alpha=1}^{d_A} \left\vert \psi^{\alpha k} \right\vert^2 
  ~.
\end{align*}
We see that $\braket{v_j}{v_k}$ is a Gramian matrix of vectors $\ket{v_j} \in
\mathcal{H}_A$ that conveys the information about the reduced state $\rho^B$
on the subspace $\mathcal{H}_A$. Though the vectors
$\ket{v_k}$ are not normalized, we readily define their normalized counterpart:
\begin{align*}
\ket{\chi_k} & \coloneqq \frac{\ket{v_k}}{\sqrt{\braket{v_k}{v_k}}} \\
  & = \sum_{\alpha=1}^{d_A} \frac{\psi^{\alpha k}}{\sqrt{\sum_{\beta=1}^{d_A} \left\vert \psi^{\beta k} \right\vert^2 }} \ket{a_\alpha}
  ~.
\end{align*}
And, eventually, we obtain:
\begin{align}
\rho^A = \sum_{k=1}^{d_B} p_k^A \ketbra{\chi_k^A}{\chi_k^A}
  ~,
\label{eq:PartialA}
\end{align}
where $\left\{\ket{\chi_j}\right\}_{j=1}^{d_B}$ is a set
of $d_B$ pure states on $\mathcal{H}_A$ which, usually,
are nonorthogonal. This provides the following geometric quantum 
state, at fixed $d_B$:
\begin{align*}
  \mu^{A}_{d_B}(dZ)
  & \coloneqq  \sum_{k=1}^{d_B} p_k^B \delta_{\chi_k}  \left(dZ\right)
  ~,
\end{align*}
where $\delta_{\chi_k}$ is the Dirac measure with support only on
the point $\chi_k \in \mathcal{P}(\mathcal{H}_A)$ corresponding to the ket $\ket{\chi_k}$.

While it is possible to track all information about
$\left\{p_k^A\right\}_{k=1}^{d_B}$ for small $d_B$, in the thermodynamic limit
this rapidly becomes infeasible. A probabilistic description becomes more
appropriate. One could object that this is not a concern since, at each step in
the limit, the spectral decomposition $\rho^A = \sum_{i=1}^{d_A} \lambda_i
\ket{\lambda_i}\bra{\lambda_i}$, where the $\lambda_i$ are the Schmidt
coefficients of $\ket{\psi_{AB}}$, is always available. However, this retains
only $\rho^A$'s matrix elements, erasing the information contained in the
vectors $\ket{v_j} = \sqrt{p^A_j} \ket{\chi^A_j}$. That is, $\rho^B$ has been
erased from the description.

However, this information can be crucial to understanding $A$'s behavior. The
geometric formalism resolves this issue as it naturally keeps the ``relevant''
information by handling measures and probability distributions. In the limit of
a large ``environment'' $B$, despite the fact that storing all information
about the environment's details is exponential in $B$'s size, the geometric
quantum state's form (convex sum of Dirac deltas) facilitates working with
smooth approximations of increasing accuracy. It does so by retaining the
information about its ``purifying environment''. 

Since we are interested here in the thermodynamics, one needs to operationally 
define the thermodynamic-limit procedure. We do so by confining ourselves to
modular systems and defining an iterative procedure. Modular systems are those
made by identical subsystems, each described by a Hilbert space $\mathcal{H}_d$
of dimension $d$. Thus, we imagine our system to contain $N_A$ such repetitive
units, while the environment contains $N_B \geq N_A$. This means $\mathcal{H}_A
= \mathcal{H}_d^{\otimes N_A}$ and $\mathcal{H}_B = \mathcal{H}_d^{\otimes
N_B}$, so that $d_A = d^{N_A}$ and $d_B = d^{N_B}$. At any given iteration, the
joint system will always be in a pure state $\ket{\psi_{AB} (N_B)} \in
\mathcal{H}_A \otimes \mathcal{H}_B$.

We also imagine that the system's global dynamics has a Hamiltonian $H_{AB}$ of
fixed functional form. For example, the XXZ model. Starting with $N_B = N_A$,
at each step we add one repetitive unit $N_B \to N_B + 1$ and choose a series
of pure states $\left\{ \ket{\psi_{AB}(N_B)}\right\}_{N_B}$ with the required
property that the limit of the average energy has to be finite:
\begin{align*}
\lim_{N_B \to \infty} \frac{\bra{\psi_{AB}(N_B)}H_{AB} \ket{\psi_{AB}(N_B)}}{N_A + N_B} = \varepsilon
   ~.
\end{align*}

For example, one can decide to consistently pick the ground state of the
Hamiltonian $H_{AB}$. In general, though, there is no unique way of performing
the procedure. However, with any specific choice of the series
$\left\{\ket{\psi_{AB}(N_B)}\right\}_{N_B}$ satisfying the constraint on
average energy, the procedure is well-defined, physical, and meaningful. It
provides an operational way to study the thermodynamic limit of the geometric
quantum state $\mu^A_{d_B}$.

That said, by no means does this guarantee the limit always exists. However, it
does allow exploring it in a physically meaningful way. In particular, given
this operational implementation of the thermodynamic limit, we say that:
\begin{align*}
\lim_{d_B \to \infty} \mu_{d_B}^A = \mu^A_{\infty}
  ~,
\end{align*}
This requires a geometric quantum state $\mu_{\infty}^A$ on
$\mathcal{P}(\mathcal{H}_A)$ such that, for any $\epsilon > 0$ arbitrarily
small, one can always find some finite $\overline{d}_B$ such that for any $d_B
\geq \overline{d}_B$ one has that $D(\mu_{d_B}^A,\mu_{\infty}^A) \leq
\epsilon$. Here, $D(\mu,\nu)$ is a notion of distance between geometric quantum
states that we take to be the measure-theoretic counterpart of the total
variation distance: $D(\mu,\nu)\coloneqq \sup_{S \in \mathcal{B}}\left\vert
\mu(S) - \nu(S)\right\vert$, where $\mathcal{B}$ is $\sigma$-algebra of
$\mathcal{P}(\mathcal{H})$'s Borel sets.

When the limit exists, we say that the thermodynamic limit of the geometric
quantum state is $\mu_{\infty}^A$ or, equivalently, $P^A_{\infty}$:
\begin{align*}
P_{\infty}^A\left[ \mathcal{O}\right]
  & = \int_{\mathcal{P}(\mathcal{H}_A)} \!\!\!\!\!\!\!\!\!\!\!\!
  \mu_{\infty}^A(dZ) \mathcal{O}(Z) \\
  & = \sum_{k=1}^{\infty} p_k^A \mathcal{O}(\chi_k^A)
  ~.
\end{align*}

$P^A_{\infty}$ is a functional whose operational meaning is understood in 
terms of ensemble theory, as explained above. Geometric quantum states 
describe ensembles of independent and noninteracting instances of the same 
quantum system whose pure states are distributed according to a given probability 
distribution. Loosely speaking, if we pick a random pure state out of the ensemble 
described by $P^A_{\infty}$, the probability of finding it in a small region around 
$Z$ is $dP_Z = \mu_{\infty}^A( dZ )$. 

\section{From geometry to statistics}
\label{sec:GeoStats}

Several observations serve to motivate defining statistical mechanics
using the geometric formalism. Consider a large system consisting of a
macroscopic number $M$ of qubits from which we extract, one by one,
$\mathcal{N}$ qubit states. Describing small subsystems of a macroscopic 
quantum system places us in the realm of quantum statistical mechanics. 
It is therefore reasonable to assume that the qubit states are distributed
according to Gibbs' canonical state $\gamma_\beta = e^{-\beta H}/Z_\beta$.
This is statistically meaningful by means of ensemble theory and, thus,
interpreted as a collection of identical noninteracting systems, each in an
energy eigenstate, with relative frequency given by Boltzmann rule.  

However, one can see how the assumption that all systems must be in
one of the energy eigenstates can be relaxed. After we extract the
$k$-th sample from the macroscopic system, that sample's state is supposed to
be an energy eigenstate $\ket{E_i^{(k)}}$ with probability $p(Z(\vert E_i^{(k)}
\rangle)) \propto e^{-\beta E_i^{(k)}}$. \emph{A priori}, however, there is no 
reason to assume that the Hamiltonians $H_k$ of all the samples are identical 
to each other. In fact, $\ket{E_i^h}\neq \ket{E_i^k}$ and $E_i^h \neq E_i^k$.
Even if they are, in principle there is no reason why the qubits should be in 
their energy eigenstates. This point was originally made by Khinchin \cite{Khin51} and Schr\"odinger
\cite{Sch52}, who advocated for the use of ensembles of wave-functions.

To address this, a description of the system's state that does not contain this 
assumption is provided by the continuous counterpart of Gibbs canonical state,
first introduced in Ref. \cite{Brody1998}, written as the following functional:
\begin{align*}
P_\beta [A] = \frac{1}{Q_\beta[h]}\intP  e^{-\beta h(Z)} A(Z)dV_{FS}
  ~,
\end{align*}
where:
\begin{align*}
Q_\beta[h] = \intP e^{-\beta h(Z)} dV_{FS}
  ~,
\end{align*}
with $h(Z) = \sum_{\alpha \beta} H_{\alpha \beta} Z^\beta \overline{Z}^\alpha$.
While this distribution retains a characteristic feature of the canonical Gibbs 
ensemble:
\begin{align*}
\frac{p_\beta \big(Z(\ket{E_n}) \big)}{p_\beta \big(Z(\ket{E_m})\big)}
  = e^{-\beta(E_n - E_m)}
  ~,
\end{align*}
it also extends this ``Boltzmann'' rule to arbitrary states:
\begin{align*}
- \log  \left[\frac{p_\beta \big(Z(\ket{\psi}) \big) }{p_\beta
  \big(Z(\ket{\phi})\big)} \right]
  = \beta \left[ h(Z(\psi))-h(Z(\phi)) \right]
  ~.
\end{align*}

Therefore, formulating the statistical mechanics of quantum states via the
geometric formalism differs from the standard development, based on an
algebraic formalism. This becomes obvious when we write the Gibbs canonical
density matrix $\gamma_\beta$ in the geometric formalism:
\begin{align*}
p_{\mathrm{Gibbs}}(Z)
  & = \sum_{k=0}^{D-1} \frac{e^{-\beta E_k}}{\Tr e^{-\beta H}}
  \delta [Z - Z(\ket{E_k})] \\
  & \neq \frac{e^{-\beta h(Z)}}{Q_\beta[h]}
  ~.
\end{align*}
This makes explicit the standard formalism's assumption that the measure is
Dirac-like---peaked on energy eigenstates.


Despite quantum statistical mechanics' undeniable successes, this assumption is
not, in general, justified. In point of fact, it is the origin of the missing
environmental information noted above. These arguments motivate an alternative
formulation of the statistical mechanics of quantum systems, first introduced
in Ref. \cite{Brody1998}---one based on geometric quantum states rather than on
the familiar density matrices.

\section{Statistical treatment of geometric quantum mechanics}
\label{sec:Stat_GQM}

Representing a quantum system's state as a continuous mixed state was first
broached, to our knowledge, by Brody and Hughston \cite{Brody1998,Brody2007}.
Our goal here is to advance the idea, going from statistical mechanics to
thermodynamics. To set the stage for a \emph{geometric quantum thermodynamics},
the following first presents our version of their results, derived via the
formalism defined in Sec. \ref{sec:Extension}, and then expands on them. We
begin with the fundamental postulate of classical statistical mechanics and its
adaptation to quantum mechanics---the microcanonical and canonical ensembles.

\subsection{Classical microcanonical ensemble: A priori equal probability}

At its most basic level, the fundamental postulate of classical statistical
mechanics is that, in an isolated system's phase space, \emph{microstates
with equal energy have the same chance of being populated}. Calling $\vec{q}$
and $\vec{p}$ generalized velocities and positions, which provide a coordinate
frame for the classical phase-space, the postulate corresponds to assuming that
the \emph{microcanonical} probability distribution $P_{\mathrm{mc}}$ of finding
the system in a microstate $(\vec{p},\vec{q})$ is, at equilibrium:
\begin{align*}
P_{\mathrm{mc}}(\vec{q},\vec{p})
  = \begin{cases}
  1 / W(\mathcal{E}) & \mathrm{if~}
  E(\vec{q},\vec{p}) \in [\mathcal{E},\mathcal{E} + \delta \mathcal{E}] \\
  0 & \mathrm{otherwise}
  \end{cases}
  ~.
\end{align*}
Here, $W(\mathcal{E})$ is the number of microstates $(\vec{q},\vec{p})$
belonging to energy shell $I_{\mathrm{mc}} \coloneqq [\mathcal{E},\mathcal{E}+\delta \mathcal{E}]$:
\begin{align*}
W(\mathcal{E}) = \int_{E(\vec{q},\vec{p}) \in I_{\mathrm{mc}}} d\vec{q} \wedge d \vec{p} 
  ~,
\end{align*}
with $\int d\vec{q} \wedge d \vec{p} ~P_{\mathrm{mc}}(\vec{q},\vec{p}) = 1$.

\subsection{Quantum microcanonical ensemble: A priori equal probability}

Quantum statistical mechanics relies on the quantum version of the Gibbs
ensemble. For macroscopic isolated systems this is usually interpreted as the
quantum system having equal chance $p_{\mathrm{mc}}$ to be in any one of the energy eigenstates $\ket{E_n}$, as long as $E_n \in I_{\mathrm{mc}}$:
\begin{align*}
p_{\mathrm{mc}}(E_n) = \begin{cases}
  1/{\mathrm{W_{\mathrm{mc}}}} & \mathrm{if} 
  E_n \in [\mathcal{E},\mathcal{E} + \delta \mathcal{E}] \\
  0 & \mathrm{otherwise}
  \end{cases}
  ~.
\end{align*}
Here, $W_{\mathrm{mc}} = \sum_{E_n \in I_\mathrm{mc}} 1$ is the number of
energy eigenstates that belong to the microcanonical window $I_{\mathrm{mc}}$.
Thus, the equal-probability postulate provides the following definition for
the microcanonical density matrix:
\begin{align*}
\rho_{\mathrm{mc}} = \frac{1}{W_{\mathrm{mc}}} \sum_{E_n \in I_{\mathrm{mc}}} \ket{E_n}\bra{E_n}
  ~.
\end{align*}
Geometric quantum mechanics gives an alternative way to extend
equal-probability to quantum systems, which we discuss now.

\subsection{Geometric quantum microcanonical ensemble: A priori equal probability}

The following summarizes an approach to the statistical mechanics of quantum
systems first presented in Refs. \cite{Brody1998,Brody2007,Brody2007b}. In geometric quantum mechanics the role of the Hamiltonian operator as the generator of unitary dynamics is played by the real quadratic function:
\begin{align*}
h(Z) = \sum_{\alpha \beta}H_{\alpha \beta}Z^\alpha \overline{Z}^\beta
  ~,
\end{align*}
where $H_{\alpha \beta}$ are the matrix elements of the Hamiltonian operator in
a reference basis; see Eq. (\ref{eq:LiouvilleDynamics}). As $h$ is the
generator of Liouville dynamics on the pure-state manifold
$\mathcal{P}(\mathcal{H})$, it is easy to see that there is a straightforward
geometric implementation of the a-priori-equal-probability postulate in the
quantum setting:
\begin{align*}
p_{\mathrm{mc}}(Z) = \begin{cases}
	1 / \Omega(\mathcal{E}) &
	h(Z) \in I_{\mathrm{mc}},
	~\text{for~all~} Z \in \mathcal{P}(\mathcal{H}) \\
	0 & \mathrm{otherwise}
	\end{cases}
  ~.
\end{align*}
Due to normalization, $\Omega(\mathcal{E})$ is the volume of the quantum-state manifold enclosed by the microcanonical energy shell $I_{\mathrm{mc}}$:
\begin{align*}
\Omega(\mathcal{E}) = \int_{h(Z)\in I_{\mathrm{mc}}}  dV_{FS}
  ~.
\end{align*}
where $dV_{FS}$ is the Fubini-Study volume element introduced in Sec.
\ref{sec:GQM}. In probability-and-phase coordinate $Z^\alpha =
\sqrt{p_\alpha}e^{i\nu_\alpha}$ the volume element has the explicit form:
\begin{align*}
dV_{FS} & = \prod_{\alpha=1}^n \frac{dp_\alpha d\nu_\alpha}{2}
  ~.
\end{align*}

Following Heslot \cite{Heslot1985}, we introduce dimensional coordinates via:
\begin{align*}
Z^\alpha = \frac{X^\alpha + i Y^\alpha}{\sqrt{\hbar}}
  ~,
\end{align*}
where $X^\alpha$ and $Y^\alpha$ are real numbers with dimensions $\left[ X
\right] = \left[ \sqrt{\hbar}\right] = \mathrm{Length}
\sqrt{\mathrm{Mass}/\mathrm{Time}}$ and $\left[ Y \right] = \left[
\sqrt{\hbar}\right] = \mathrm{Momentum} \sqrt{\mathrm{Time}/\mathrm{Mass}}$.
The ratio $X/Y$ is a pure number, while their product $XY$ has the dimension
$\hbar$ of an action. Note that $dp_\alpha d\nu_\alpha / 2 = dX_\alpha
dY_\alpha / \hbar$. This allows us to write the Fubini-Study measure in a
classical fashion:
\begin{align*}
dV_{FS} & = \prod_{\alpha=1}^{D-1} \frac{dX^\alpha dY^\alpha}{\hbar} \\
  & = \frac{d\vec{X} d\vec{Y}}{\hbar^{D-1}}
  ~,
\end{align*}
where the $X^\alpha$ play the role of generalized coordinates and $Y^\alpha$
that of generalized momenta. However, it is worth noting that the global
geometry of the classical phase-space differs substantially from that of
$\mathcal{P}(\mathcal{H})$. 

Given these definitions, it is now possible to calculate the number of states
$\Omega(\mathcal{E}) \approx \omega(\mathcal{E}) \delta \mathcal{E}$, where
$\delta \mathcal{E}$ is the size of the microcanonical energy shell and
$\omega(\mathcal{E})$ is the density of states:
\begin{align*}
\omega(\mathcal{E}) & = \int_{h(Z)=\mathcal{E}} dV_{FS} \\
  & = \frac{\pi^{D-1}}{(D-1)!}  \sum_{k=0}^{D-1}
  \prod_{j \neq k, j=0}^{D-1} \frac{(\mathcal{E}-E_k)_{+}}{(E_j-E_k)}
  ~,
\end{align*}
where $(x)_{+} \coloneqq \max (0,x)$.
Since $\mathcal{E} \in [E_0,E_{\mathrm{max}}]$, there exists an $\overline{n}$
such that $\mathcal{E} \in ] E_{\overline{n}},E_{\overline{n}+1} [$. This means
that we can stop the sum at $k=\overline{n}(\mathcal{E})$ since for all $k >
\overline{n}$ we have $(\mathcal{E}-E_k)_+ = 0$. This gives:
\begin{align}
\omega(\mathcal{E}) =
  \frac{\pi^{D-1}}{(D-1)!}
  \sum_{k=0}^{\overline{n}(\mathcal{E})}
  \frac{(D-1)(\mathcal{E}-E_k)^{D-2}}{\prod_{j \neq k, j=0}^{D-1} (E_j-E_k)}
\label{eq:mc_density}
  ~.
\end{align}
This is in agreement with Ref. \cite{Brody2007}'s Eq. (5). Appendix
\ref{app:App} provides a detailed proof, using a convenient mathematical result
by Ref. \cite{Lasserre2015}.

\begin{figure}[h]
\centering
\includegraphics[width=\columnwidth]{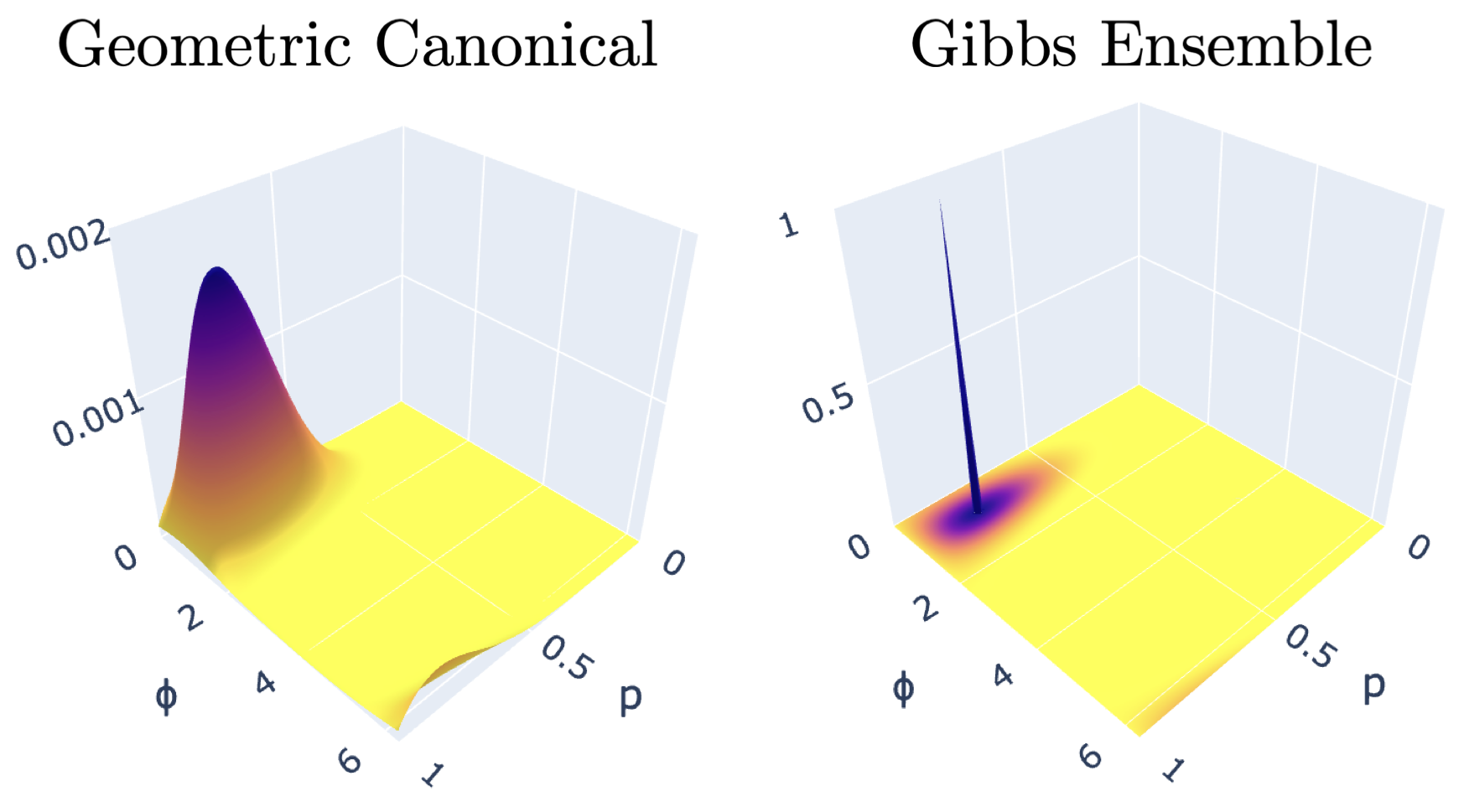}
\caption{Alternate ensembles in the geometric and standard settings:
	Differences are plainly evident. Canonical probability distributions on a
	qubit's state manifold $\mathbb{C}P^1$ with coordinates $Z = (Z^0,Z^1) =
	(\sqrt{1-q},\sqrt{q}e^{i\chi})$ where $q \in [0,1]$ and $\chi \in
	[-\pi,\pi]$. $\mathbb{C}P^1$ discretized using a $100$-by-$100$ grid on the
	$(q,\chi)$ coordinates exploiting the fact that, with these coordinates,
	the Fubini-Study measure is directly proportional to the Cartesian volume
	element $dV_{FS} = dq d\chi / 2$. The Hamiltonian is $H = \sigma_x +
	\sigma_y + \sigma_z$, with $\hbar=1$ and inverse temperature $\beta = 5$
	($\kB = 1$). (Right) Gibbs ensemble, where the measure is concentrated around
	coordinates of the respective eigenvectors
	$\big(q(\ket{E_0}),\chi(\ket{E_0})\big) = (0.789,-2.356)$ and
	$\big(q(\ket{E_1}),\chi(\ket{E_1})\big) = (0.211,0.785)$.
	(Left) Geometric Canonical Ensemble. Notice the difference in scale, due to the
	fact that the Geometric Canonical Ensemble has continuous support on the 
	quantum state space, not just on single points (energy eigenstates).
	}
\label{fig:Comparison} 
\end{figure}

\subsection{Quantum canonical ensemble: Statistical physics of quantum states}

The geometric approach to microcanonical ensembles extends straightforwardly
to the canonical case, defining the continuous canonical ensemble as:
\begin{align}
p_{\beta}(Z) = \frac{e^{-\beta h(Z)}}{Q_\beta[h]}
  ~,\label{eq:Canonical}
\end{align}
where:
\begin{align*}
Q_\beta[h] = \intP e^{-\beta h(Z)}  dV_{FS} 
  ~.
\end{align*}

Reference \cite{Brody1998} first proposed the general form of the canonical 
partition function $Q_\beta[h]$, working it out explicitly in several
low-dimensional cases. Follow-on work provided an exact formula valid for
arbitrary finite-dimensional Hilbert spaces \cite{Brody2007}. Appendix
\ref{app:App} provides an alternative proof and explicit examples of:
\begin{align}
Q_\beta[h] = \sum_{k=0}^{D-1}
  \frac{ e^{-\beta E_k} }{\prod_{j=0, j \neq k}^n (\beta E_k - \beta E_j)}
  ~.
\label{eq:GeoPartFunc}
\end{align}
This is in full agreement with Ref. \cite{Brody2007}'s Eq. (6). 

It is important to stress here that such Geometric Canonical Ensemble is 
genuinely different than the ensemble defining Gibbs' canonical density matrix. 
A visual comparision of the two ensembles, with a simple hamiltonian $\sigma_x+\sigma_y+\sigma_z$
and inverse temperature $\beta=5$ is given in Figure \ref{fig:Comparison}.

With the ensembles laid out we can now see the emergence of geometric quantum
thermodynamics, with its fundamental laws.

\section{Geometric Quantum Thermodynamics}\label{sec:GQT}

With a consistent statistical geometric quantum mechanics in hand, we can now
formulate geometric quantum thermodynamics. This is modeled via the geometric
canonical state Eq. (\ref{eq:Canonical}). Notice that, in this setting, an
appropriate entropy definition has yet to be given. Paralleling early work by
Gibbs, one can consider the functional:
\begin{align*}
H_q\left[ p\right] = - \kB \intP p(Z) \log p(Z) dV_{FS}
  ~.
\end{align*}
An information-theoretic analysis of this quantity and its relation with the
von Neumann entropy was done in Ref. \cite{Brody2000}. This functional allows
properly evaluating $p(Z)$'s entropy if and only if the dimension of the
support of $p$ has the same real dimension of $\CP{n}$. Reference
\cite{Anza21a} defined and explored the appropriate generalization to
geometric quantum states with generic support, including fractal
distributions.

Let's consider $H_q$'s role, though, for the quantum foundations of
thermodynamics. For Eq. (\ref{eq:Canonical})'s geometric canonical ensemble
this gives:
\begin{align*}
H_q & = \beta (U - F) ~,
\end{align*}
where:
\begin{align*}
U & \coloneqq \intP p_\beta(Z) h(Z)dV_{FS} \quad \text{and}\\
F & \coloneqq -\frac{1}{\beta}\log Q_\beta
\end{align*}
are, respectively, the average energy and the free energy arising from the
geometric partition function $Q_\beta$.

This means that we can directly import a series of fundamental results from
classical thermodynamics and statistical mechanics into the quantum setting,
fully amortizing the effort invested to develop the geometric formalism.

\subsection{First Law}

The first result is a straightforward derivation of the First Law:
\begin{align}
dU & = \intP dV_{FS} p(Z) dh(Z) + \intP dV_{FS} dp(Z)h(Z) \nonumber \\
  & = dW + dQ 
  ~.
  \label{eq:FirstLaw}
\end{align}
We call the contribution $dW$ \emph{work}, since it arises from a change
in the Hamiltonian $h(Z)$ generated by an external control operating on the
system. We call the contribution $dQ$ \emph{heat}, as it is associated
with a change in entropy. Indeed, by direct computation one sees that:
\begin{align*}
dH_q  = \beta dQ \qquad \text{and} \qquad dF = dW
  ~.
\end{align*}
This gives the standard form of the First Law for isothermal, quasi-static
processes:
\begin{align*}
dU = TdH_q + dF
  ~,
\end{align*}
where $T \coloneqq \left( \kB \beta \right)^{-1}$. Conforming to the
conventional statistical approach to thermodynamics, beyond energy
conservation, one can use the First Law to extract phenomenological relations
(e.g., Maxwell's relation) that hold at thermodynamic equilibrium:
$\partial U / \partial H_q = T$. In this, the partial derivatives are
intended as infinitesimal changes occurring while maintaining the system at
thermal equilibrium.

\subsection{Second Law}

The Second Law follows from the Crooks \cite{Crooks1999a} and Jarzynski
\cite{Jarzynski2011} fluctuation theorems
\cite{Campisi,Klages2013,Deffner2019}. Their treatment can be straightforwardly
exploited, thanks to the Hamiltonian nature of Schr\"odinger's equation when
written on the quantum-state manifold $\PH$.

As summarized in Eq. (\ref{eq:LiouvilleDynamics}), given a Hamiltonian
$h(Z,\lambda)$ on $\PH$ that depends on an externally-controlled parameter
$\lambda = \lambda(t)$, the unitary evolution is given by the Liouville
equation Eq. (\ref{eq:LiouvilleDynamics}) as in classical mechanics:
\begin{align*}
\frac{\partial p(Z)}{\partial t} = \left\{p(Z) , h(Z,\lambda) \right\}
  ~.
\end{align*}

Notably, one can apply Jarzynski's original argument \cite{Jarzynski1997} to
driven quantum systems, without the need to exploit the \emph{two-times
measurement scheme} \cite{Deffner2019}. The setup is standard.

The ensemble of quantum systems starts in a geometric canonical state defined
by Eq. (\ref{eq:Canonical}) and is then driven with a Hamiltonian that depends
on a parameter $\lambda$ following the time-dependent protocol
$\lambda=\lambda(t)$ with $t \in [0,1]$. This leads directly to an ensemble of
protocol realizations. That said, we define the \emph{single-trajectory work}
as:
\begin{align*}
W = \int_{0}^{1} \dot{\lambda}(t) \frac{\partial h}{\partial \lambda}
 \big( Z(\psi_t),\lambda(t) \big)dt
  ~,
\end{align*}
where $\dot{\lambda} = d \lambda / dt$ and $Z(\psi_t)$ are the homogeneous
coordinates on $\mathbb{C}P^{D-1}$ for $\ket{\psi_t}$. Therefore,
$\ket{\psi_t}$ are the solutions of Eq. (\ref{eq:unitary_evolution}).

With these premises, Jarzynski's original argument applies \emph{mutatis
mutandis} to give:
\begin{align}
\MV{e^{-\beta W}}_{\mathrm{ens}}
  & = \frac{Q_\beta [h(\lambda_f)]}{Q_\beta [h(\lambda_i)]} \nonumber \\
  & = e^{-\Delta F}
  ~,
    \label{eq:Jarzynski}
\end{align}
where $\lambda(0) = \lambda_i$ and $\lambda(1)=\lambda_f$ and
$\MV{x}_{\mathrm{ens}}$ denotes the ensemble average over many protocol
realizations. From this, one directly applies Jensen's inequality:
\begin{align*}
\MV{e^{-\beta W}}_\text{ens}
   \geq e^{-\beta \MV{W}}_\text{ens}
\end{align*}
to obtain the Second Law's familiar form:
\begin{align}
\MV{W}_\text{ens} \geq F
  ~.
\label{eq:SecondLaw}
\end{align}

\section{Geometric thermalization in a phenomenological model}\label{sec:QubitsExample}

The validity of geometric quantum thermodynamics, as defined above, hinges on
the assumption of (geometric) thermal equilibrium. It therefore implicitly
relies on a dynamical mechanism driving the system towards the geometric
canonical ensemble. This section shows that this occurs in at least one model
for the out-of-equilibrium dynamics of a single qubit. 

A quantum system interacting with its surroundings evolves in a nonunitary
fashion due to the fact that it exchanges energy (or other extensive
quantities) and so becomes correlated with its environment. This can be
modeled using the theory of open quantum systems and its dissipative dynamics
\cite{Bre02,Weiss12,Gar10,Carm93}. While most approaches focus on establishing
an equation governing the dynamical evolution of the system's density matrix,
here we are interested in the thermodynamics of the geometric quantum state as
the ensemble behind the density matrix. A principled description and modeling
of the dynamics of an open quantum system within the geometric approach is
beyond the present scope. Though, its development is currently ongoing.

Instead, the following shows how to represent dissipation within the geometric
formalism for a stochastic model. This serves a twofold purpose. First, it
provides simple examples of how geometric quantum mechanics evolves open
quantum systems in a variety of cases. Second, it supports the theory
developed above with a numeric analysis of an experimentally relevant
scenario.

While the emphasis is still on the geometric formalism, and its natural
phase-space geometry, this approach is not far from ``Stochastic Schrodinger
Equations''. See, for example, Refs. \cite{Bou04,Bre02,Bel89b,Bel89a,Bel88})
that import techniques from the classical theory of stochastic processes. The
following exploits this idea, applying it to the geometric language and
drawing from a variety of known approaches. It does so by examining a
phenomenological model for dissipative dynamics that, as we show, exhibits
thermalization towards the geometric canonical ensemble. 

It considers the stochastic dynamics of a two-level system with state space
$\PH \sim \mathbb{C}P^{1}$. Generally, this results from a two-state
approximation of a more complex system interacting with an environment. It
gives a standard approximation that provides sensible results in a variety of
physical regimes. These include systems that inherently consist of two states,
such as spin-$1/2$, chiral molecules
\cite{Urra12,Barg11,Rod14,Rod13,Barg13,Barg14,Katz16}, and atoms at low
temperature, considering only the two lowest states.  They also include,
though, continuous-variable systems in a double-well potential, Josephson
junctions \cite{Feyn63c}, and effective descriptions of macroscopic
condensates. As a related technical aside beyond quantum mechanics, we note
that the proper analysis and simulation of stochastic dynamics on Riemannian
manifolds is a topic of its own interest \cite{Gom21,Rob60}.

Accounting for the nonisolated nature of the system involves modeling the
environment and the latter's effect on the effective qubit. This, therefore,
depends on the specific case under study and leads to different effective
equations governing the qubit's nonequilibrium behavior. From the system's
perspective, however, a general setup is available in a regime in which
coupling with the environment is weak and the environment is effectively large
and disordered. These approximations are expected to hold for large
environments, where one can argue for the emergence of stochastic dynamics for
the evolution of the open system.

The prototypical case, in which a specific form of these equations can be
derived by integrating out the environmental degrees of freedom, is given by
the Caldeira-Leggett model \cite{Cald83,Cald83b,Legg87} with an environment
modeled by an infinite number of noninteracting harmonic oscillators.
Respecting these approximations' validity, a generic model of Langevin-like
dynamics on $\mathbb{C}P^1$ is:
\begin{align}
\dot{p} & = -\partial_\phi E + V_p + W_p~,\\
\dot{\phi} & =  \partial_p E + V_\phi + W_{\phi}
  ~, \nonumber
\end{align}
in $(p,\phi)$ coordinates. In this, $E=E(p,\phi)$ is an effective Hamiltonian
generating the deterministic part of the dynamics; see Eq.
(\ref{eq:unitary_evolution}). This is a renormalized version of the system's
Hamiltonian. $V_{p}$ and $V_{\phi}$ depend linearly on $(p,\phi)$ and
$(\dot{p},\dot{\phi})$. They describe (i) dissipative mechanisms such as
friction, modeled with a dependence on $\dot{p}$ or $\dot{\phi}$, as in
standard Langevin equations, and (ii) unstable states, modeled with a
dependence $V_p = -kp$ to allow for exponential decay $p_{\mathrm{decay}}(t)
\sim p_0e^{-kt}$, as in a two-level atom decaying into its ground state.

Finally, $W_p$ and $W_{\phi}$ are stochastic variables with no drift that
account for the environment's mixing effect on the system. When the
environment is sufficiently large and unstructured, they can be modeled as
Gaussian processes, $\mathbb{E}[W_a(s+t)W_b(s)]=\mathbb{E}[W_a(t)W_b(0)]
\approx \delta_{ab} \gamma_a \delta(t)$, with $a,b \in \left\{ p,\phi\right\}$
and $\gamma_a \propto \kB T$, with $T$ the temperature of the environment.
This is true in the Caldeira-Leggett model for Ohmic baths.

As anticipated above, specific forms of these equations have successfully
modeled the evolution of a variety of two-level systems. We also note how, in
several cases, and also in Refs. \cite{Mill78,Mey79,Kostin72,Stock97,Chru04},
this approach to open quantum systems is quite similar to GQM as it relies on
canonical representations of the quantum state space. For chiral molecules,
for example, one has $E(p,\phi)=\delta \MV{\sigma_x} + \epsilon
\MV{\sigma_z}=\delta 2 \sqrt{p(1-p)}\cos \phi + \epsilon (1-2p)$, $V_p = - k
\dot{p}$, with $k \sim 10^{-1}$, $W_\phi=V_\phi=0$ and $W_p(t)$ white noise
with strength $\gamma_p \propto \kB T$. The thermodynamics arising from this
set of dynamical equations has been studied in detail
\cite{Urra12,Barg11,Rod14,Rod13,Barg13,Barg14,Katz16}. 

The goal here is rather to showcase the experimental relevance of the
geometric canonical ensemble. The following does so showing, numerically, that
the evolution provided by the stochastic equations above leads to the
dynamical emergence of the geometric canonical ensemble. This is directly
relevant to the out-of-equilibrium dynamics of an ensemble of chiral molecules
or of an ensemble of experiments with Josephson junctions.

The specific stochastic equations under study are:
\begin{align}
\dot{p} & =  \delta 2 \sqrt{p(1-p)} \sin \phi - k_d p - k_f \dot{\phi}
  + \sqrt{\gamma} \xi(t) \label{eq:stoch_model} \\
\dot{\phi} & =  -\delta \frac{1-2p}{\sqrt{p(1-p)}} \cos \phi
  + 2\epsilon
  \nonumber 
  ~,
\end{align}
where $k_d$ and $k_f$ are coefficients accounting for dissipation mechanisms,
such as instability of a state and friction. Up to simple re-definition of
variables, that does not change the physics, the model with $k_d=0$ is the
same as in Ref. \cite{Rod14,Rod13}.

Exploiting the Markovian character of Gaussian noise, the statistics of many
independent realizations of this stochastic process on $\CP{1}$ can be
extracted by examining the time-aggregated statistics of a single, very long,
trajectory. We thus simulate the long-time dynamics of a qubit initiated in a
fully out-of-equilibrium configuration $q_0(p,\phi) = \delta (p-p_0) \delta
(\phi-\phi_0)$, corresponding to a pure state
$\ket{p_0,\phi_0}=\sqrt{1-p_0}\ket{0} + \sqrt{p_0}e^{i\phi_0}\ket{1}$, where
$\ket{0},\ket{1}$ are the standard computational basis. For chiral molecules,
these are the (left and right) chiral eigenstates. Here, we show the results
for $p_0 = 0.9$ and $\phi=4 \pi /3$ and checked that they do not depend on
this choice. Results are shown for parameter values $\delta = \epsilon = 1$,
$\gamma = 0.2$, and $k_d=0$. While these match the model in Ref.
\cite{Rod14,Rod13}, the results are largely independent of this specific
choice and hold for broad regimes in $(\delta,\epsilon,k_d,\gamma)$ parameter
space. 

The analysis was performed as follows. After generating a single long-time
trajectory using the Milstein method, we collected statistics
$\tilde{P}_{nk}$. We then generated a histogram to approximate the probability
that, at any given time, the system is found in a small region of the state
space $\tilde{P}_{nk} \approx \overline{\mathrm{Pr}}\left[Z \in
\mathcal{I}_{nk}\right]=\lim_{T \to \infty} \int_0^{T} \int_{\mathcal{I}_{nk}}
q_t(Z) dV_{FS}$. In this, $\left\{\mathcal{I}_{nk}\right\}_{n,k=1}^{N}$ is a
coarse graining of $\CP{1}$ in which each region $\mathcal{I}_{nk} =
[p_n,p_{n+1}] \times [\phi_k, \phi_{k+1}]$ has the same Fubini-Study volume
$\mu_{FS}(\mathcal{I}_{nk})=N^{-2}$, $p_k = n/N$, and
$\phi_k = 2\pi k/N$. Reference \cite{Anza21a} gives a detailed
analysis of why this is an appropriate coarse-graining, its
information-theoretic relevance, and how to generalize it to arbitrary $\CP{n}$.

Concretely, the numerical analysis used $N=50$. The dynamics was generated
setting $T=10^2$ in units in which $\hbar=\delta=1$. This was chosen by
numerically checking that the reconstructed histogram does not change
significantly when increasing $T$. The time window $[0,T]$ was discretized to
use the Milstein algorithm to generate Gaussian noise with $dt = 10^{-4}$.
These, again, are consistent with the choices in Refs. \cite{Rod14,Rod13}. In
short, the number of time steps $N_T = 10^6$, with $N_T dt = T$.

\begin{figure}[h]
\centering
\includegraphics[width=\columnwidth]{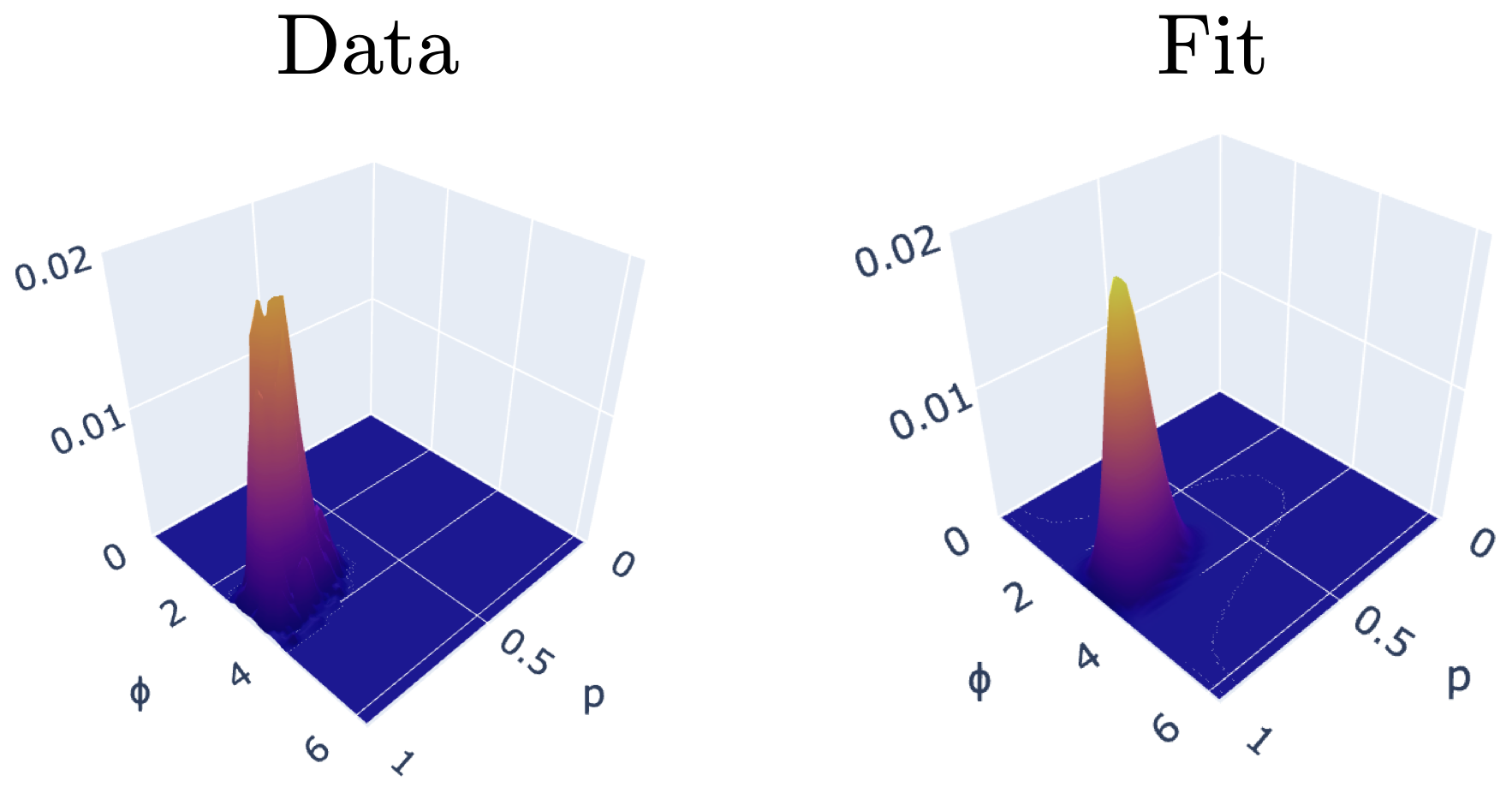}
\caption{Comparing time-aggregated data of a single trajectory
	generated by Eq. (\ref{eq:stoch_model})'s stochastic model (left)
	to the fit to a geometric canonical ensemble with functional form as in
	Eq. (\ref{eq:Canonical}) (right). Here, $h(Z) = E(p,\phi) =\delta
	\MV{\sigma_x} + \epsilon \MV{\sigma_z}=\delta 2 \sqrt{p(1-p)}\cos \phi +
	\epsilon (1-2p)$, with $\delta=\epsilon=1$. The excellent agreement is
	visually clear, and it is quantified by a total variation distance between the
	two distributions of $f \approx 5.6  \times 10^{-4}$.
	}
\label{fig:DataFitComparison} 
\end{figure}

To extract the inverse temperature $\beta$ the collected statistics were used
to perform a $2$D least-square fit to the geometric canonical ensemble. The
latter's appropriateness was established by using the following figure of
merit: $f = \sum_{n,k} |\tilde{P}_{nk} - q_{nk}^{\mathrm{fit}} |^2 \in [0,1]$,
where $q_{nk}^{\mathrm{fit}} = Q^{-1}{\beta^*} \int_{\mathcal{I}_{nk}}
dV_{FS}e^{-\beta^{*} E(Z)}$ and $\beta^{*}$ is the optimal value extracted
from the least-square fit. This is the total variation distance between the
coarse-grained geometric quantum states obtained from the data
$\left\{\tilde{P}_{nk}\right\}_{n,k}$ and the one obtained from the best fit
to the geometric canonical ensemble
$\left\{q_{nk}^{\mathrm{fit}}\right\}_{n,k}$. It ranges from zero to one and
is the classical analog of the well-known trace-distance for density matrices.
At selected parameters, $f\approx 5.6 \times 10^{-4}$. This quantifies the
visually excellent agreement seen in Fig. \ref{fig:DataFitComparison}.

Before drawing broad conclusions, a few comments are in order regarding
specific results. First, thermalization is observed even when changing
parameter values. This is true for \emph{any} of the Hamiltonian parameters,
$\delta$ and $\epsilon$. Moreover, there are good numerical indications that
this holds for any $k_d > 0$. However, $k_d$ and $\gamma$ do affect the
effective (inverse) temperature $\beta^{*}$ the system reaches. Analyzing how
this happens and the underlying mechanisms is beyond the present scope, which
aimed only at establishing the predictive relevance of the geometric canonical
ensemble in an experimentally realistic settings.

Second, we ignored issues related to the time-scale at which the aggregated
geometric quantum state reaches the canonical form. These were bypassed by
using a time window $[0,T]$ that guaranteed the aggregated data does not
change when increasing $T$.

Third, Eq. (\ref{eq:stoch_model})'s model arises from a bath that is a set of
noninteracting harmonic oscillators with Ohmic correlation functions and
interactions linear in the phase difference $\phi$, leading to a friction
$\propto \dot{\phi}$. A different kind of interaction is possible, linear in
the population $p$, that leads to a friction term $\dot{p}$. While not
reported here, there are numerical indications that this alternative exhibits
thermalization to the geometric canonical ensemble as well. This supports the
intuition that thermalization is mostly driven by the lack of memory of the
stochastic term, ultimately due to the Ohmic nature of the bath's correlation
functions. 

Fourth, the effective nature of the description makes the model widely
applicable. And so, a number of straightforward generalizations would be quite
interesting to explore. These include, for example, changing the noise
structure to accommodate limited memory and allowing for competition between
the different ways in which the system interacts with the harmonic bath and
the decay in both $p$ and $\phi$. Of particular interest, both conceptually
and practically, is determining which terms lead to dynamical localization and
what kinds of system-bath interactions are necessary for these terms to emerge
by integrating out the bath degrees of freedom.

\section{Summary and Conclusion}\label{sec:Discussion}

While historically quantum mechanics is firmly rooted in an algebraic
formalism, an alternative based on the differential geometry of quantum state
space $\mathcal{P}(\mathcal{H}) \sim \mathbb{C}P^{D-1}$ is readily available.

As previous works repeatedly emphasized
\cite{Bengtsson2017,STROCCHI1966,Heslot1985}, the geometric approach brings
quantum and classical mechanics much closer, aiming to leverage the best of
both. The space $\PH$ of quantum states is a K\"ahler space, with two
intertwined notions of geometry---Riemannian and symplectic. It also sports a
preferred notion of measure, selected by invariance under unitary
transformations---the Fubini-Study measure. One can exploit this rich
geometric structure to define generic probability measures on $\PH$. The
result is a new kind of quantum state---the \emph{geometric quantum
state} \cite{Anza20a}---that generalizes the familiar density matrix but
provides more information about a quantum system's physical configuration.
Essentially, it expresses the multitude of ensembles, induced by different
environments, behind a density matrix.

Leveraging parallels between the geometric formalism and classical mechanics,
the statistical treatment of geometric quantum mechanics provides a continuous
counterpart of Gibbs ensembles. Section~\ref{sec:GQT} laid out how to
establish quantum thermodynamics on the basis of the geometric formalism.
Building on Section~\ref{sec:Stat_GQM}'s statistical treatment of geometric
quantum mechanics, it derived the First and Second Laws of Geometric Quantum
Thermodynamics. Despite the two results appearing identical to the existing
laws, derived within standard quantum statistical mechanics, they involve
quantities that are genuinely different. Understanding how Eqs.
(\ref{eq:FirstLaw}), (\ref{eq:Jarzynski}), and (\ref{eq:SecondLaw}) connect to
their standard counterparts \cite{Deffner2019} is a challenge that we must
leave for the future. We note Ref. \cite{Campisi13} obtained a similar result
that, lacking the geometric perspective, considered microcanonical and
canonical ensembles of pure states, as first advocated by Khinchin
\cite{Khin51} and Schr\"odinger \cite{Sch52}.

Remarkably, predictions from standard quantum statistical mechanics and its
geometric counterpart differ. This poses a challenge: Which theory should one
use? Ultimately, this problem does not have a generic solution. Answering the
question requires understanding the details of the long-time dynamic of an
open quantum system and, in general, this will be be model-specific. Here, to
showcase the relevance of the geometric approach, we showed that there is a
class of known stochastic models, aimed at describing chiral molecules and
Josephson's junctions, that indeed does exhibit dynamical evolution towards
the geometric canonical ensemble. One thus expects the predictions from
geometric quantum thermodynamics to hold in the cases where the dynamical
model in Eq. (\ref{eq:stoch_model}) is justified.

The geometric approach to quantum thermodynamics opens the door to new and
interesting questions and novel research avenues. Let's mention two. First, the
ensemble interpretation of geometric quantum mechanics suggests employing the
geometric formalism to describe the thermodynamics of ensembles, rather than
relying on that of density matrices. The main advantage is that this
delineates the environmental resources required to support a given density
matrix. Indeed, while two different experimental setups can give rise to the
same density matrix, their difference implicitly lies in the distinct ways the
density matrix is created. This is directly relevant to the energetics of
information processing technologies built from quantum computers and quantum
sensors.

Second, from a conceptual perspective, geometric quantum thermodynamics and
statistical mechanics are as at least as powerful as their standard
counterpart. Yet, they can make different predictions. Self-consistency of
thermodynamic predictions suggests that this difference should be negligible
in a truly macroscopic regime in which both system and environment are
macroscopically large. This is, however, a highly nontrivial statement whose
proof requires a much better understanding of the emergence of thermodynamic
predictions from fully dynamical considerations. We believe the new research
avenues, together with the larger perspective provided by geometric quantum
mechanics, will greatly enrich our understanding of the phenomenology of
many-body quantum systems.

\section*{Acknowledgments}
\label{sec:acknowledgments}

F.A. thanks Anthony Ciavarella, Marina Radulaski, Davide Pastorello, and
Davide Girolami for discussions on the geometric formalism of quantum
mechanics. F.A. and J.P.C. thank David Gier, Dhruva Karkada, Samuel Loomis,
and Ariadna Venegas-Li for helpful discussions and the Telluride Science
Research Center for its hospitality during visits. This material is based upon
work supported by, or in part by, Templeton World Charity Foundation, Inc. grant TWCF0336, 
FQXi Grant FQXi-RFP-IPW-1902, and U.S. Army Research Laboratory and the U. S. Army 
Research Office under contracts W911NF-18-1-0028 and W911NF-21-1-0048.

\section*{Data Availability Statement}
The data that support the findings of this study are available from the
corresponding authors upon reasonable request.

%
%
%

\appendix

\section{Independent result}
\label{sec:Lasserre}

For completeness, the following summarizes Ref. \cite{Lasserre2015}'s result
called on in calculating the density of states. Given the $n$-simplex
$\Delta_n: \left\{ \vec{x} \in \mathbb{R}^{n}_{+} \,\, : \vec{e} \cdot \vec{x}
\leq 1\right\}$, where $\vec{e}$ is the vector of ones in $\mathbb{R}^{n}$, a
section of the simplex is defined by a vector $\vec{a}\in \mathbb{S}^{n}$ and
we want to compute the $n$-dimensional and $(n-1)$-dimensional volume of the
following sets:
\begin{align*}
\Theta(\vec{a},t) & \coloneqq \Delta_n \cap
  \left\{ \vec{x} \in \mathbb{R}^{n} : \vec{a}^T \cdot \vec{x} \leq t\right\}
  ~\text{and} \\
S(\vec{a},t) & \coloneqq \Delta_n \cap
  \left\{ \vec{x} \in \mathbb{R}^{n} : \vec{a}^T \cdot \vec{x} = t\right\}
  ~,
\end{align*}
where $\vec{a}^T$ is the transpose of $\vec{a}$. The result assumes flat
geometry, which is obtained from the volume element $dp_1dp_2\ldots dp_n$.
Letting $(x)_{+}\coloneqq \max (0,x)$ and $a_0=0$, then:
\begin{align*}
\mathrm{Vol}\left( \Theta(\vec{a},t)\right)
  & = \frac{1}{n!} \sum_{k=0}^{n} \frac{(t-a_k)_{+}^{n}}{\prod_{j \neq k \,,\,
  j=0 }^n (a_j - a_k)} \\
  & = \frac{1}{n!} \frac{t^{n}}{\prod_{k=1}^{n} a_k} + \frac{1}{n!} \sum_{k =1}^{n} \frac{(t-a_j)_+^{n}}{\prod_{j \neq k \, , \, j=0}^n (a_j-a_k)}
\end{align*}
and:
\begin{align*}
\mathrm{Vol}\left( S(\vec{a},t)\right)
  & = \frac{1}{(n-1)!} \sum_{k=0}^{n} \frac{(t-a_k)_{+}^{n-1}}{\prod_{j \neq k
  \,,\, j=0 }^n (a_j - a_k)} \\
  & = \frac{1}{(n-1)!} \frac{t^{n-1}}{\prod_{k=1}^{n} a_k} \\
  & \quad + \frac{1}{(n-1)!} \sum_{k =1}^{n} \frac{(t-a_j)_+^{n-1}}{\prod_{j \neq k \, , \, j=0}^n (a_j-a_k)}
  ~.
\end{align*}

\section{Geometric Quantum Density of States and Canonical Ensemble}

Again for completeness, we first recall the basic definitions, given in the
main text, used in the two sections that follow to calculate the density of
states and statistical physics of quantum states in the geometric formalism.

\subsection{Setup and notation}

Consider a Hilbert space $\mathcal{H}$ of finite-dimension $D$. The manifold
$\mathcal{P}(\mathcal{H})$ of states is the complex projective space
$\mathbb{C}P^{D-1}$. A point $Z$ on the manifold is a set of $D$ homogeneous and
complex coordinates $\{ Z^\alpha \}$. A point corresponds to a pure state with the
identification $Z \leftrightarrow \ket{\psi} = \sum_{\alpha=0}^{D-1}Z^\alpha
\ket{e_\alpha}$, where $\left\{\ket{e_\alpha}\right\}_\alpha$ is an arbitrary
but fixed basis of $\mathcal{H}$. This parametrization underlies the choice of
a reference basis that, however, is ultimately irrelevant. While concrete
calculations of experimentally measurable quantities can be made easier or
harder by an appropriate coordinate system, the overall result is independent
on such choices. The quantum mechanical expectation value is a quadratic and
real function on the manifold of the quantum states:
\begin{align*}
a(Z) & \coloneqq \bra{\psi(Z)}A\ket{\psi(Z)} \\
  & = \sum_{\alpha,\beta=0}^{D-1} A_{\alpha,\beta}Z^\alpha \overline{Z}^\beta
  ~.
\end{align*}
When $A=H$ is the system's Hamiltonian, the function $a(Z) = h(Z)$ generates the
vector field $V_H$ on $\mathbb{C}P^{D-1}$. The associated Hamiltonian equations
of motion become the Schr\"odinger equation (and its complex conjugate) when
using the standard formalism with Hilbert spaces. In the geometric formalism,
states are functionals from the algebra of observables to the real numbers.
Effectively, they are probability distributions, both discrete and continuous,
on the quantum-state manifold $\mathbb{C}P^{D-1}$.

\subsection{Microcanonical density of states: Proof of Eq. (\ref{eq:mc_density})}

We start with the a priori equal probability postulate and build the microcanonical shell as follows:
\begin{align*}
p_{\mathrm{mc}}(Z) = 
  \begin{cases}
  1 / W(\mathcal{E}) & \mathrm{if}
  ~h(Z) \in [\mathcal{E},\mathcal{E} + \delta \mathcal{E}] \\
  0 & \mathrm{otherwise}
  \end{cases}
  ~.
\end{align*}
Due to normalization we have:
\begin{align*}
W(\mathcal{E}) = \int_{h(z)\in I_{\mathrm{mc}}}  dV_{FS}
  ~,
\end{align*}
where $dV_{FS}$ is the volume element of the Fubini-Study metric:
\begin{align*}
dV_{FS} = \frac{1}{2^n} dp_1 dp_2 \ldots dp_n d\nu_1 \ldots d\nu_n
  ~.
\end{align*}
This gives the manifold volume:
\begin{align*}
\mathrm{Vol}(\mathbb{C}P^n) = \frac{\pi^n}{n!}
  ~.
\end{align*}
For concrete calculations, normalize the measure so that $\mathbb{C}P^{D-1}$'s
total volume is unity, using:
\begin{align*}
d \mu_{n} & = \frac{dV_{FS}}{\mathrm{Vol}(\mathbb{C}P^n)} \\
  & = \frac{n!}{(2\pi)^n} \prod_{k=1}^n dp_k \prod_{k=1}^n d\nu_k 
  ~.
\end{align*}
This does not alter results in the main text. On the one hand, calculations of
measurable quantities are independent of this value. On the other, here, at the
calculation's end, we reintroduce the appropriate normalization.

We can now compute $W(\mathcal{E})$ for a generic quantum system. Assuming that
$\delta \mathcal{E} \ll \vert E_{\mathrm{max}}- E_{\mathrm{min}}\vert$, we have
$W(\mathcal{E}) = \Omega(\mathcal{E}) \delta \mathcal{E}$ and $\Omega(\mathcal{E})$ is
the area of the surface $\Sigma$ defined by $h(Z) = \mathcal{E}$:
\begin{align*}
\Omega(\mathcal{E}) = \int_\Sigma d \sigma
  ~,
\end{align*}
where $d\sigma$ is the area element resulting from projecting both the
symplectic two-form and the metric tensor onto the surface $\Sigma$. To compute
this we choose an appropriate coordinate system:
\begin{align*}
Z^\alpha & = \braket{E_\alpha}{\psi(Z)} \\
  & = n_\alpha e^{i\nu_\alpha}
\end{align*}
adapted to the surface $\Sigma$:
\begin{align*}
h(Z) & = \bra{\psi(Z)} H \ket{\psi(Z)} \\
  & = \sum_{k=0}^n E_k |\braket{\psi}{E_k}|^2 \\
  & = \sum_{k=0}^n E_k n^2_k \\
  & = \mathcal{E}
  ~.
\end{align*}

On both sides we subtract the ground state energy $E_0$ and divide by $E_{\mathrm{max}} - E_0$ to obtain the following defining equation for $\Sigma \subset \mathbb{C}P^n$:
\begin{align*}
F(n_0,n_1,\ldots, n_n, \nu_1, \ldots, \nu_n)
  & = \sum_{k=0}^n \varepsilon_k n_k^2 - \varepsilon \\
  & = 0 
  ~,
\end{align*}
with:
\begin{align*}
 \varepsilon_k & = \frac{E_k - E_0}{E_{\mathrm{max}}-E_0} \in [0,1]
 ~\text{and}\\
  \varepsilon & = \frac{\mathcal{E}-E_0}{E_{\mathrm{max}}-E_0} \in [0,1]
  ~.
\end{align*}
We use octant coordinates for $\mathbb{C}P^n$:
\begin{align*}
(Z_0,Z_1,\ldots,Z_n)
  = \left( n_0,n_1e^{i\nu_1},n_2e^{i\nu_2} \ldots, n_n e^{i\nu_n}\right)
  ~,
\end{align*}
where $n_k \in [0,1]$ and $\nu_k \in [0,2\pi[$.
With the transformation $p_k = n_k^2$ the equation for $\Sigma$ becomes:
\begin{align*}
\sum_{k=0}^n p_k \varepsilon_k - \varepsilon = 0
  ~.
\end{align*}

\subsubsection{Qubit Case}

The state space of a single qubit is $\mathbb{C}P^1$. The latter's
parametrization:
\begin{align*}
p \varepsilon_0 + (1-p) \varepsilon_1 = 1-p
\end{align*}
means that $h(Z) \leq \mathcal{E}$ is equivalent to $1-p \leq \varepsilon$.
The volume is therefore given by:
\begin{align*}
\mathrm{Vol}_{n=1}(\mathcal{E})
  & = \frac{1}{\pi} \int_{h(\phi)\leq \mathcal{E}} dV_{FS} \\
  & = \frac{1}{2\pi} \int_{1-\varepsilon}^1 dp \int_{0}^{2\pi} d \nu \\
  & = \varepsilon \\
  & = \frac{\mathcal{E}-E_0}{E_1-E_0}
  ~.
\end{align*}
In turn, this gives;
\begin{align*}
W_{n=1}(\mathcal{E}) 
  & = \mathrm{Vol}_{n=1}(\mathcal{E}+\delta \mathcal{E}) -
  \mathrm{Vol}_{n=1}(\mathcal{E}) \\
  & = \frac{1}{E_1-E_0} \delta \mathcal{E}
  ~.
\end{align*}
In other words:
\begin{align*}
\Omega_{n=1}(\mathcal{E}) = \frac{1}{E_1-E_0}
  ~,
\end{align*}
which is a constant density of states.

\subsubsection{Qutrit Case}

The state space of qutrits is $\mathbb{C}P^2$, with parametrization $Z =
(Z_0,Z_1,Z_2) = (1-p-q,p e^{i\nu_1},q e^{i\nu_2})$. With these coordinates, the
equations defining the constant-energy hypersurface is:
\begin{align*}
(1-p-q) \varepsilon_0 + p \varepsilon_1 + q \varepsilon_2
  = p \varepsilon_1 + q \leq \varepsilon
  ~.
\end{align*}
And, it has volume:
\begin{align*}
\mathrm{Vol}_{n=2}(\mathcal{E})
  & = \frac{2}{(2\pi)^2} \int \!\!\! \int dq dq \int \!\!\! \int d\nu_1 d\nu_2 \\
  & = 2 \int \!\! \! \int_S dp dq
  ~.
\end{align*}
In this, we have the surface $S \coloneqq \left\{ (p,q) \in \mathbb{R}^2 \,\,:
p,q \geq 0 , p+q \leq 1, q \leq \varepsilon - p \varepsilon_1\right\}$.
Examining the geometry we directly see that the region's area is:
\begin{align*}
A(S) =
  \begin{cases}
  \frac{1}{2} - \frac{1}{2} \frac{(1-\varepsilon)^2}{1-\varepsilon_1}
  & \mathrm{when~} \varepsilon \geq \varepsilon_1 \\
  \frac{\varepsilon^2}{2\varepsilon_1}
  & \mathrm{when~} \varepsilon < \varepsilon_1
  \end{cases}
 ~.
\end{align*}
Or:
\begin{align*}
A(S) =
  \begin{cases}
  \frac{1}{2} - \frac{1}{2} \frac{(E_2 - \mathcal{E})^2}{(E_2-E_1)(E_2-E_0)}
  & \mathrm{when ~} \mathcal{E} \geq E_1 \\
  \frac{1}{2}\frac{(\mathcal{E}-E_0)^2}{(E_1-E_0)(E_2-E_0)}
  & \mathrm{when~} \mathcal{E} < E_1
  \end{cases}
  ~.
\end{align*}
One can check that the function $A(S)[\mathcal{E}]$ and its first derivative
are continuous. Eventually, we have:
\begin{align*}
W_{n=2}(\mathcal{E})
  & = \mathrm{Vol}_{n=2}(\mathcal{E}+\delta \mathcal{E})
  - \mathrm{Vol}_{n=2}(\mathcal{E}) \\
  & = \begin{cases}
  \frac{2(E_2 - \mathcal{E})}{(E_2-E_1)(E_2-E_0)} \delta \mathcal{E}
  & \mathrm{when~} \mathcal{E} \geq E_1 \\
  \frac{2(\mathcal{E} - E_0)}{(E_2-E_0)(E_1-E_0)} \delta \mathcal{E}
  & \mathrm{when~} \mathcal{E} < E_1
  \end{cases}
  ~.
\end{align*}

\subsubsection{Generic Qudit Case: $\mathbb{C}P^n$}

To use Ref. \cite{Lasserre2015}'s result, summarized in App. \ref{sec:Lasserre},
we must change coordinates. Again, using ``probability + phase'' coordinates:
\begin{align*}
& \sum_{k=0}^n p_k E_k = \mathcal{E}
\end{align*}
means that:
\begin{align*}
\sum_{k=1}^n p_k a_k & = t(\mathcal{E})  \\
  a_k & = a(E_k) \\
  & = \frac{E_k-E_0}{R} ~, \\
  R & = \sqrt{\sum_{k=1}^n \left( E_k - E_0\right)^2} ,~\text{and} \\
  t(\mathcal{E}) & = \frac{\mathcal{E}-E_0}{R}
  ~.
\end{align*}
In this way, we can apply the result, finding:
\begin{align*}
\mathrm{Vol}_n \left( \mathcal{E}\right)
  & = \sum_{k=0}^{n}
  \frac{(t-a_k)_{+}^{n}}{\prod_{j \neq k \,,\, j=0 }^n (a_j - a_k)} \\
  & =\sum_{k=0}^n
  \frac{(\mathcal{E}-E_k)_{+}^n}{\prod_{j \neq k, j=0}^n (E_j-E_k)}
  ~.
\end{align*}
Since $\mathcal{E} \in [E_0,E_{\mathrm{max}}]$, there exists an $\overline{n}$
such that $\mathcal{E} \in ]E_{\overline{n}},E_{\overline{n}+1}[$. This means
that the sum in the second term stops at $k=\overline{n}$ because after that
$(\mathcal{E}-E_k)_+ = 0$. Hence, there exists $\overline{n}(\mathcal{E})$ such
that for all $k > \overline{n}$ we have $(\mathcal{E}-E_k)_+ = 0$. This, in
turns, shows that:
\begin{align*}
\mathrm{Vol}_n \left( \mathcal{E}\right)
  = \sum_{k=0}^{\overline{n}(\mathcal{E})}
 \frac{(\mathcal{E}-E_k)^n}{\prod_{j \neq k, j=0}^n (E_j-E_k)}
  ~.
\end{align*}
This leads to the desired fraction of $\mathbb{C}P^n$ microstates in a microcanonical energy shell $[\mathcal{E},\mathcal{E}+ d \mathcal{E}]$:
\begin{align*}
W_n(\mathcal{E})
 & = \Omega_n(\mathcal{E}) d \mathcal{E} \\
 & = \left(  \sum_{k=0}^{\overline{n}(\mathcal{E})}
 \frac{n(\mathcal{E}-E_k)^{n-1}}{\prod_{j \neq k, j=0}^n (E_j-E_k)}\right)
 d \mathcal{E} 
  ~.
\end{align*}
This allows defining the \emph{statistical entropy} $S(\mathcal{E})$ of a
quantum system with finite-dimensional Hilbert space of dimension $D=n+1$ as:
\begin{align*}
S(\mathcal{E}) = \log W_{D-1}(\mathcal{E}) 
  ~.
\end{align*}

\subsection{Statistical physics of quantum states: Canonical ensemble}
\label{app:App}

The continuous canonical ensemble is defined as:
\begin{align*}
\rho_{\beta}(\psi) = \frac{e^{-\beta h(\psi)}}{Q_\beta[h]}
  ~,
\end{align*}
where:
\begin{align*}
Q_\beta[h] = \int_{\mathbb{C}P^{D-1}} e^{-\beta h(\psi)}  dV_{FS} 
  ~.
\end{align*}
The following analyzes the simple qubit case and then moves to the general
treatment of a finite-dimensional Hilbert space $\mathcal{H}$.

\subsubsection{Single Qubit}

The Hilbert space here is $\mathcal{H}$ while the pure-state manifold is
$\mathbb{C}P^1$. And so, we have:
\begin{align*}
Q_\beta[h] = \frac{1}{4} \int_0^{\pi} d \theta \sin \theta
  \int_{0}^{2\pi} d\phi \,\, e^{-\beta h(\theta,\phi)}
 ~,
\end{align*}
with $h(\theta,\phi) = \vec{\gamma} \cdot \MV{\vec{\sigma}} = \vec{\gamma}
\cdot \vec{b}(\theta,\phi)$.

Since we consider a single qubit, whose state space is $S^2$ embedded in
$\mathbb{R}^3$, we can write $\vec{\gamma} \cdot \vec{b}(\psi) = |\!|
\vec{\gamma} |\!| \cos \theta$, where $\theta$ is the angle between
$\vec{\gamma}$ and $\vec{b}(\psi)$. Thus, we can use an appropriate coordinate
$h(\phi,\theta) = |\!| \vec{\gamma} |\!| \cos \theta$ aligned with
$\vec{\gamma}$ to find:
\begin{align*}
Q_\beta[h]
  = \pi \frac{\sinh \beta |\!|\vec{\gamma}|\!|}{\beta |\!|\vec{\gamma}|\!|}
  ~.
\end{align*}
Or, using ``probability + phase'' coordinates $(p,\nu)$ we can also write:
\begin{align*}
\frac{1}{2}\int_0^1 dp \int_0^{2\pi} d \nu \,\,e^{-\beta [(1-p)E_0 + pE_1]} = \pi \frac{e^{-\beta E_0} - e^{-\beta E_1}}{\beta(E_1-E_0)}
  ~.
\end{align*}
The change in coordinates is given by the result of diagonalization: $E_{0} = -
|\!|\vec{\gamma}|\!|$ and $E_{1} = |\!|\vec{\gamma}|\!|$. This
yields the expected result:
\begin{align*}
Q_\beta[h]
  & = \pi \frac{e^{-\beta E_0} - e^{-\beta E_1}}{\beta(E_1-E_0)} \\
  & = \pi \frac{\sinh \beta |\!|\vec{\gamma}|\!|}{\beta |\!|\vec{\gamma}|\!|}
  ~.
\end{align*}

\subsubsection{Generic Treatment of $\mathbb{C}P^n$}

We are now ready to address the general case of qudits:
\begin{align*}
Q_\beta[h]
  & = \int_{\mathbb{C}P^n} e^{-\beta h(Z)} dV_{FS} \\
  & = \frac{1}{2^n} \int \prod_{k=0}^n e^{-\beta p_k E_k}
  \prod_{k=1}^n dp_k d\nu_k \\
  & = \pi^n \int_{\Delta_n} \prod_{k=0}^n e^{-\beta p_k E_k}
  \delta \left(\sum_{k=0}^np_k-1\right) dp_1 \ldots dp_n
  ~.
\end{align*}
To evaluate the integral we first take the Laplace transform:
\begin{align*}
I_n(r) \coloneqq \int_{\Delta_n} \prod_{k=0}^n e^{-\beta p_k E_k}
  \delta \left(\sum_{k=0}^np_k-r\right) dp_1 \ldots dp_n
\end{align*}
to get:
\begin{align*}
\tilde{I}_n(z) \coloneqq \int_0^{\infty} e^{-zr}I(r) dr
  ~.
\end{align*}
Calculating, we find:
\begin{align*}
\tilde{I}_n(z)
  & = \prod_{k=0}^n \frac{(-1)^k}{(\beta E_k + z)} \\
  & = (-1)^{\frac{n(n+1)}{2}} \prod_{k=0}^n \frac{1}{z-z_k}
  ~.
\end{align*}
with $z_k = - \beta E_k \in \mathbb{R}$.

The function $\tilde{I}_n(z)$ has $n+1$ real and distinct poles: $z=z_k =
-\beta E_k$. Hence, we can exploit the partial fraction decomposition of
$\tilde{I}_n(z)$, which is:
\begin{align*}
(-1)^{\frac{n(n+1)}{2}} \prod_{k=0}^n \frac{1}{z-z_k}
  = (-1)^{\frac{n(n+1)}{2}} \sum_{k=0}^n \frac{R_k}{z-z_k}
  ~,
\end{align*}
where:
\begin{align*}
R_k & = \left[ (z-z_k)\tilde{I}_n(z)\right]_{z=z_k} \\
  & = \prod_{j=0, \, j \neq k}^n\frac{(-1)^{\frac{n(n+1)}{2}}}{z_k-z_j}
  ~.
\end{align*}

The inverse Laplace transform's linearity, coupled with the basic result:
\begin{align*}
\mathcal{L}^{-1}\left[\frac{1}{s+a}\right](t) = e^{-at}\Theta(t)
  ~,
\end{align*}
where:
\begin{align*}
\Theta(t) = \begin{cases}
  1 & t \geq 0 \\
  0 & t < 0
  \end{cases}
  ~,
\end{align*}
gives:
\begin{align*}
I_n(r) & = \mathcal{L}^{-1}[\tilde{I}_n(z)](r) \\
  & = \Theta(r) \sum_{k=0}^n R_k e^{z_k r}
  ~.
\end{align*}
And so, we finally see that:
\begin{align*}
Q_\beta[h] & = I_n(1) \\
  & = \sum_{k=0}^n
  \frac{ e^{-\beta E_k} }{\prod_{j=0, \,\, j \neq k}^n (\beta E_k - \beta E_j)}
  ~.
\end{align*}

\bibliography{chaos,library}

\end{document}